\documentclass[10pt]{article}
\usepackage{multicol}
\usepackage{graphicx}
\usepackage{amsmath}
\usepackage[a4paper]{geometry}
\usepackage{enumerate}
\usepackage{hyperref}

\begin{document}

\begin{center}
{\Large \bf Kinetic freeze-out temperature and transverse flow
velocity\\ in Au-Au collisions at RHIC-BES energies}

\vskip1.0cm

Muhammad~Waqas{\footnote{E-mail: waqas\_phy313@yahoo.com}},
Bao-Chun~Li{\footnote{Corresponding author. E-mail:
libc2010@163.com; s6109@sxu.edu.cn}}
\\

{\small\it Institute of Theoretical Physics \& Department of
Physics \&\\ State Key Laboratory of Quantum Optics and Quantum
Optics Devices,\\ Shanxi University, Taiyuan, Shanxi 030006,
China}

\end{center}

\vskip1.0cm

{\bf Abstract:} Based on the data-driven analysis, the
mid-rapidity transverse momentum ($p_T$) spectra of charged
hadrons ($\pi^+$, $K^+$ and $p$) produced in central and
peripheral gold-gold (Au-Au) collisions from the Beam Energy Scan
(BES) program at the Relativistic Heavy Ion Collider (RHIC) are
fitted by the blast-wave model with Boltzmann-Gibbs statistics.
The model result are in agreement with the experimental data
measured by the STAR Collaboration at the RHIC-BES energies. We
observe that the kinetic freeze-out temperature ($T_0$),
transverse flow velocity ($\beta_T$), mean transverse momentum
($\langle p_T\rangle$), and initial temperature ($T_i$) increase
with the collision energy and with the event centrality.
\\



\vskip1.0cm
\begin{multicols}{2}

{\section{Introduction}}

One of the most fundamental questions in nuclear matter is to
determine the phase structure of the strongly-interacting quantum
chromodynamics (QCD) matter~\cite{1a,1b,1c}. The yield ratios,
transverse momentum ($p_T$) spectra and other data for various
identified particles produced in proton-proton ($pp$),
proton-nucleus ($pA$) and nucleus-nucleus ($AA$) collisions at
high energies are important observable quantities for determining
the phase structure. The experimental facilities, for example the
Relativistic Heavy-Ion Collider (RHIC) and the Large Hadron
Collider (LHC) provide excellent tools to study the properties of
Quark-Gluon Plasma (QGP)~\cite{1d,1e,1f}.

The phase diagram of the QCD matter is usually expressed in terms
of the chemical freeze-out temperature ($T_{ch}$) and the baryon
chemical potential ($\mu_B$)~\cite{1g,1h}. Besides, other
quantities such as the kinetic freeze-out temperature ($T_{kin}$
or $T_0$) and transverse flow velocity ($\beta_T$) are useful to
understand the phase diagram~\cite{1i}. To search for the possible
critical energy in the phase transition from hadronic matter to
QGP in high energy collisions, the STAR Collaboration has been
performing the Beam Energy Scan (BES) program~\cite{1j,1k,1L,1m}
at the RHIC. Besides, other experiments at similar or lower
energies at other accelerators are scheduled~\cite{1n,1o}.

Generally, the processes of high energy collisions result
possibly in three main stages~\cite{1p,1q,1}:

\begin{enumerate}[i)]

\item The initial stage: at this stage the collisions are in the
beginning. The temperature at this stage is called the initial
temperature which is one of the main factors to affect the
particle spectra, which is less studied in the community
comparatively. After the initial state, the ``fireball" leads to a
decrease in the temperature and finally to the hadronization.

\item The chemical freeze-out stage: at this stage the inner
collisions among various particles are elastic and the yield
ratios of differential types of particles remain invariant. The
chemical freeze-out temperature $T_{ch}$ can be obtained from the
particle ratios, which is much studied in the community
comparatively.

\item The kinetic freeze-out stage: at this stage the scattering
processes stop and the hadrons decouple from the rest of the
system and the hadron's energy/momentum spectra freeze in time.
The temperature at this stage is known as the kinetic freeze-out
temperature $T_0$ which can be obtained from the $p_T$ spectra.

\end{enumerate}

When one studies $T_0$ from the $p_T$ spectra, the effect of
$\beta_T$ should be eliminated. If the effect of $\beta_T$ is not
eliminated in the temperature, this temperature is called the
effective temperature ($T_{eff}$ or $T$). At the stage of kinetic
freeze-out, $T_0$ and $\beta_T$ are two important parameters which
describe the thermal motion of the produced particles and the
collective expansion of the emission source respectively. The
spectra in low-$p_T$ region ($p_T=2$--3 GeV/$c$) which is mainly
contributed by the soft excitation process essentially separate
the contribution of the thermal motion and the collective
expansion, if one only extracts $T_0$ and $\beta_T$. The spectra
in high-$p_T$ region are contributed by the hard scattering
process which is not needed in extracting $T_0$ and $\beta_T$.

We are very interested in the extraction of $T_0$ and $\beta_T$ in
collisions at the RHIC-BES energies which are very suitable to
study the spectra in low-$p_T$ region, where the spectra in
high-$p_T$ region are not produced due to not too high energies.
In this work, the double differential $p_T$ spectra of charged
particles dependences on collision energy and event centrality in
gold-gold (Au-Au) collisions are analyzed by the blast-wave model
with Boltzmann-Gibbs statistics by means of data-driven analysis.
The model results are compared with the data measured by the STAR
Collaboration at the RHIC-BES energies~\cite{26,28}.

The remainder of this work consists of the method and formalism,
results and discussion as well as conclusions. We shall describe
the remanent parts orderly.
\\

{\section{The method and formalism}}

Various methods can be used for the extraction of $T_0$ and
$\beta_T$, e.g. the blast-wave model with Boltzmann-Gibbs
statistics~\cite{2,3,4}, the blast-wave model with Tsallis
statistics~\cite{5,5a,5b}, an alternative method by using the
Boltzmann-Gibbs statistics~\cite{3,6,7,8,9,10,11,12} and the
alternative method by using Tsallis
distribution~\cite{12,12a,12b,12c,12d,13,14}. In this work, we
choose the blast-wave model with Boltzmann-Gibbs statistics due to
its similarity with the ideal gas model in thermodynamics and few
parameters. However, these methods only describe the spectra in
low-$p_T$ region. For the spectra in high-$p_T$ region if
available, the Hagedorn function which is know as the inverse
power-law~\cite{15,16} can be used. We shall discuss these issues
in detail as follows.

In general, there are two main processes responsible in the
contribution of $p_T$ spectra. They are i) the soft excitation
process which contributes the soft component in low-$p_T$ region
and ii) the hard scattering process which contributes the hard
component in high-$p_T$ region.

For the soft component, according to refs.~\cite{2,3,4}, the
probability density function of the $p_T$ spectra in the
blast-wave model with Boltzmann-Gibbs statisitcs results in
\begin{align}
f_1(p_T)=&\frac{1}{N}\frac{dN}{dp_T} =C p_T m_T \int_0^R rdr \nonumber\\
& \times I_0 \bigg[\frac{p_T \sinh(\rho)}{T_0} \bigg] K_1
\bigg[\frac{m_T \cosh(\rho)}{T_0} \bigg],
\end{align}
where $N$ is the number of particles, $C$ is the normalization
constant, $m_T=\sqrt{p_T^2+m_0^2}$ is the transverse mass, $m_0$
is the rest mass of the considered particle, $r$ and $R$ are the
radial position and the maximum radial position respectively,
$I_0$ and $K_1$ are the modified Bessel functions of the first and
second kinds respectively, $\rho= \tanh^{-1} [\beta(r)]$ is the
boost angle, $\beta(r)= \beta_S(r/R)^{n_0}$ is a self-similar flow
profile, $\beta_S$ is the flow velocity on the surface,  and
$n_0=2$ is used in original form~\cite{2}. Particularly,
$\beta_T=(2/R^2)\int_0^R r\beta(r)dr = 2\beta_S/(n_0+2)$$
=0.5\beta_S$. The parameter $n_0$ is used different in different
works, e.g. $n_0=1$ or non-integer in refs.~\cite{5,18}, which
corresponds to the centrality from center to periphery.

Equation (1) and similar or related functions are not enough to
describe the whole $p_T$ spectra. In particular, the maximum $p_T$
reaches up to 100 GeV/$c$ in collisions at the LHC~\cite{19}.
Then, one needs other functions such as the
Tsallis--L{\'e}vy~\cite{19a,19b} or Tsallis--Pareto-type
function~\cite{19a,19c} and the Hagedorn function~\cite{15,16} or
inverse power law~\cite{23,24,25} to the spectra in high and very
high-$p_T$ regions. In this work, the hard component is simply
represented by the inverse power law. That is
\begin{align}
f_2(p_T)=\frac{1}{N}\frac{dN}{dp_T}= Ap_T \bigg( 1+\frac{p_T}{
p_0} \bigg)^{-n},
\end{align}
where $p_0$ and $n$ are free parameters and $A$ is the
normalization constant which is related to the free parameters.

However, the structure of $p_T$ spectra is very complex. In fact,
several regions have been observed and analyzed in ref.~\cite{20}.
These regions include the first one with $p_T<4$--6 GeV/$c$, the
second one with 4--6 GeV/$c$ $<p_T<17$--20 GeV/$c$ and the the
third one with $p_T>17$--20 GeV/$c$. Different regions maybe
correspond to different mechanisms. The first $p_T$ region in our
discussion is regarded as the region of soft excitation process,
while the second and third $p_T$ regions are regarded as the
regions of hard and very hard excitation process respectively. In
particular, a special region with $p_T<0.2$--0.3 GeV/$c$ is
considered due to the resonant production in some cases, and it is
regarded as the region of very soft excitation process.

Generally, all the $p_T$ regions discussed above can be unifiedly
superposed by two methods: i) the general superposition in which
the contribution regions of different components overlap each
other and ii) the Hagedorn model (the usual step
function)~\cite{15} in which there is no overlapping of different
regions of different components.

Considering $f_1(p_T)$, $f_2(p_T)$, $f_{VS}(p_T)$ and
$f_{VH}(p_T)$ which denote the probability density functions by
the soft, hard, very soft and very hard components respectively,
where $f_{VS}(p_T)$ and $f_{VH}(p_T)$ are assumed to be in the
form of $f_1(p_T)$ and $f_2(p_T)$ respectively, the unified
superposition according to the first method is
\begin{align}
f_0(p_T)=&\frac{1}{N}\frac{dN}{dp_T}=k_{VS} f_{VS}(p_T) +kf_1(p_T)\nonumber\\
&+(1-k-k_{VS}-k_{VH})f_2(p_T)\nonumber\\
&+k_{VH} f_{VH}(p_T),
\end{align}
where $k_{VS}$ is the contribution fraction of very soft
component, while $k$ and $k_{VH}$ denote the contributions of soft
and very hard components respectively.

The step function can be used to structure the superposition
according to Hagedorn model~\cite{15}, i.e.
\begin{align}
f_0(p_T)=&\frac{1}{N}\frac{dN}{dp_T}=A_{VS}\theta(p_{VS}-p_T)f_{VS}(p_T)\nonumber\\
&+A_1\theta(p_T-p_{VS})\theta(p_1-p_T)f_1(p_T)\nonumber\\
&+A_2\theta(p_T-p_1)\theta(p_{VH}-p_T)f_2(p_T)\nonumber\\
&+A_{VH}\theta(p_T-p_{VH})f_{VH}(p_T),
\end{align}
where $A_{VS}$, $A_1$, $A_2$ and $A_{VH}$ are the constants which
make the interfacing components link to each other perfectly.

Particularly, if the contributions of very soft and very hard
components can be neglected, Eqs. (3) and (4) are simplified to be
\begin{align}
f_0(p_T)=\frac{1}{N}\frac{dN}{dp_T}=kf_1(p_T)+(1-k)f_2(p_T)
\end{align}
and
\begin{align}
f_0(p_T)=&\frac{1}{N}\frac{dN}{dp_T}=A_1 \theta(p_1-p_T) f_1(p_T)\nonumber\\
&+A_2 \theta(p_T-p_1)f_2(p_T)
\end{align}
respectively. Further, if the contribution of hard component at
the RHIC-BES energies can be neglected, Eqs. (5) and (6) are
simplified to be the same form
\begin{align}
f_0(p_T)=\frac{1}{N}\frac{dN}{dp_T}=f_1(p_T).
\end{align}

This work deals with Au-Au collisions at the RHIC-BES energies,
for which Eq. (7) i.e. Eq. (1) is suitable. In the following
section, we shall use Eq. (1) to fit the experimental data
measured by the STAR Collaboration at the RHIC-BES
energies~\cite{26,28}.
\\

{\section{Results and discussion}}

Figure 1 presents the event centrality dependent double
differential $p_T$ spectra, $(1/2\pi p_T)d^2N/dp_Tdy$, of $\pi^+$,
$K^+$ and $p$ produced in the mid-rapidity interval $|y|<0.1$ in
Au-Au collisions at the center-of-mass energy per nucleon pair
$\sqrt{s_{NN}}=7.7$ GeV at the RHIC-BES, where $y$ denotes the
rapidity. The symbols represent the experimental data measured by
the STAR Collaboration~\cite{26} and the curves are our fitting
results by using the blast-wave model with Boltzmann-Gibbs
statistics, Eq. (1)~\cite{2,3,4}. The spectra in centrality class
0--5\%, 5--10\%, 10--20\%, 20--30\%, 30--40\%, 40--50\%, 50-60\%,
60--70\% and 70--80\% are scaled by 1, 1/2, 1/4, 1/6, 1/8, 1/10,
1/12, 1/14 and 1/16 respectively. The related parameters along
with $\chi^2$ and degree of freedom (dof) are listed in Table 1,
where the centrality classes are listed together. One can see that
Eq. (1) fits well the data in Au-Au collisions at 7.7 GeV at the
RHIC.

\begin{figure*}[htb!]
\begin{center}
\hskip-0.153cm
\includegraphics[width=15cm]{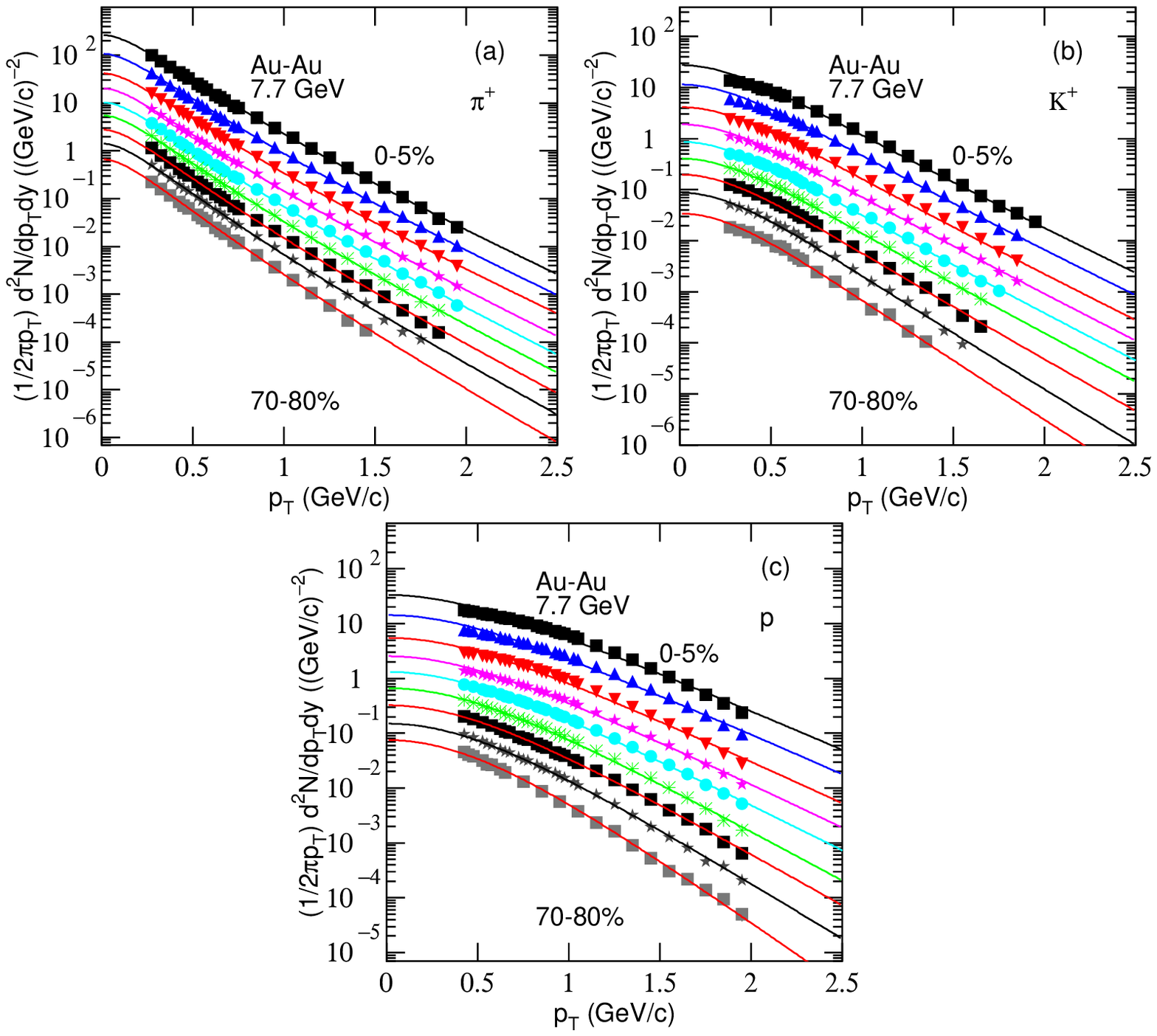}
\end{center}
Fig. 1. Transverse momentum spectra of (a)-(c) $\pi^+$, $K^+$ and
$p$ produced in different centrality bins in Au-Au collisions at
$\sqrt{s_{NN}}=7.7$ GeV. The symbols represent the experimental
data measured by the STAR Collaboration in the mid-rapidity
interval $|y|<0.1$~\cite{26}. The curves are our fitted results by
Eq. (1). The ratios of Data/Fit corresponding to the panels
(a)-(c) are presented by panels (a*)-(c*) respectively.
\end{figure*}

\begin{table*}
{\small Table 1. Values of free parameters ($T_0$ and $\beta_T$),
normalization constant ($N_0$), $\chi^2$, and dof corresponding to
the curves in Figs. 1--6. \vspace{-.50cm}
\begin{center}
\begin{tabular}{cccccccc}\\ \hline\hline
Figure & Particle & Centrality & $T_0$ & $\beta_T$ & $N_0$ &
$\chi^2$ & dof\\ \hline
Fig. 1  &$\pi^+$ & 0--5\%   & $0.130\pm0.004$ & $0.306\pm0.006$ & $15.00\pm1.00$   & 17 & 26\\
Au-Au   &        & 5--10\%  & $0.129\pm0.005$ & $0.305\pm0.005$ & $6.02\pm0.9$     & 14 & 26\\
7.7 GeV &        & 10--20\% & $0.128\pm0.004$ & $0.303\pm0.007$ & $2.35\pm1.05$    & 20 & 26\\
        &        & 20--30\% & $0.126\pm0.005$ & $0.302\pm0.006$ & $1.09\pm0.60$    & 18 & 26\\
        &        & 30--40\% & $0.124\pm0.004$ & $0.300\pm0.006$ & $0.54\pm0.43$    & 19 & 26\\
        &        & 40--50\% & $0.122\pm0.003$ & $0.297\pm0.007$ & $0.28\pm0.02$    & 13 & 26\\
        &        & 50--60\% & $0.120\pm0.004$ & $0.292\pm0.004$ & $0.14\pm0.02$    & 20 & 25\\
        &        & 60--70\% & $0.118\pm0.004$ & $0.280\pm0.005$ & $0.067\pm0.001$  & 16 & 24\\
        &        & 70--80\% & $0.115\pm0.005$ & $0.269\pm0.007$ & $0.030\pm0.008$  & 11 & 21\\
\cline{2-8}
        & $K^+$  & 0--5\%   & $0.133\pm0.005$ & $0.305\pm0.005$ & $3.65\pm0.20$    & 99 & 23\\
        &        & 5--10\%  & $0.131\pm0.004$ & $0.304\pm0.005$ & $1.50\pm0.15$    & 76 & 25\\
        &        & 10--20\% & $0.130\pm0.004$ & $0.302\pm0.007$ & $0.53\pm0.05$    & 55 & 25\\
        &        & 20--30\% & $0.128\pm0.005$ & $0.301\pm0.007$ & $0.24\pm0.17$    & 27 & 25\\
        &        & 30--40\% & $0.127\pm0.004$ & $0.299\pm0.009$ & $0.11\pm0.01$    & 27 & 24\\
        &        & 40--50\% & $0.125\pm0.006$ & $0.296\pm0.008$ & $0.050\pm0.002$  & 16 & 23\\
        &        & 50--60\% & $0.123\pm0.005$ & $0.278\pm0.007$ & $0.23\pm0.02$    & 34 & 22\\
        &        & 60--70\% & $0.118\pm0.004$ & $0.265\pm0.008$ & $0.0093\pm0.0007$& 34 & 21\\
        &        & 70--80\% & $0.116\pm0.004$ & $0.250\pm0.008$ & $0.0034\pm0.0003$& 29 & 29\\
\cline{2-8}
        & $p$    & 0--5\%   & $0.134\pm0.004$ & $0.340\pm0.005$ & $9.31\pm0.60$    & 54 & 29\\
        &        & 5--10\%  & $0.133\pm0.005$ & $0.328\pm0.006$ & $3.90\pm0.15$    & 47 & 29\\
        &        & 10--20\% & $0.132\pm0.006$ & $0.318\pm0.008$ & $1.50\pm0.20$    & 46 & 29\\
        &        & 20--30\% & $0.130\pm0.005$ & $0.310\pm0.008$ & $0.65\pm0.05$    & 28 & 29\\
        &        & 30--40\% & $0.128\pm0.004$ & $0.301\pm0.006$ & $0.32\pm0.04$    & 14 & 28\\
        &        & 40--50\% & $0.126\pm0.005$ & $0.280\pm0.007$ & $0.15\pm0.03$    & 13 & 28\\
        &        & 50--60\% & $0.124\pm0.004$ & $0.271\pm0.006$ & $0.070\pm0.008$  & 7  & 27\\
        &        & 60--70\% & $0.122\pm0.003$ & $0.250\pm0.005$ & $0.030\pm0.004$  & 8  & 28\\
        &        & 70--80\% & $0.120\pm0.006$ & $0.204\pm0.009$ & $0.013\pm0.001$  & 14 & 21\\
\cline{1-8}
Fig. 2  &$\pi^+$ & 0--5\%   & $0.132\pm0.004$ & $0.315\pm0.005$ & $19.11\pm1.60$   & 5  & 26\\
Au-Au   &        & 5--10\%  & $0.130\pm0.005$ & $0.313\pm0.005$ & $7.60\pm1.40$    & 16 & 26\\
11.5 GeV&        & 10--20\% & $0.129\pm0.004$ & $0.312\pm0.006$ & $2.90\pm0.50$    & 14 & 26\\
        &        & 20--30\% & $0.128\pm0.003$ & $0.311\pm0.007$ & $1.36\pm0.10$    & 34 & 26\\
        &        & 30--40\% & $0.127\pm0.003$ & $0.310\pm0.008$ & $0.38\pm0.05$    & 15 & 26\\
        &        & 40--50\% & $0.126\pm0.006$ & $0.307\pm0.007$ & $0.34\pm0.03$    & 16 & 26\\
        &        & 50--60\% & $0.124\pm0.004$ & $0.305\pm0.005$ & $0.16\pm0.01$    & 7  & 26\\
        &        & 60--70\% & $0.121\pm0.004$ & $0.296\pm0.006$ & $0.080\pm0.007$  & 6  & 24\\
        &        & 70--80\% & $0.119\pm0.005$ & $0.288\pm0.008$ & $0.040\pm0.005$  & 10 & 24\\
\cline{2-8}
        & $K^+$  & 0--5\%   & $0.135\pm0.005$ & $0.314\pm0.009$ & $4.23\pm0.30$    & 64 & 25\\
        &        & 5--10\%  & $0.133\pm0.004$ & $0.312\pm0.008$ & $1.72\pm0.10$    & 62 & 26\\
        &        & 10--20\% & $0.132\pm0.006$ & $0.310\pm0.010$ & $0.60\pm0.05$    & 46 & 26\\
        &        & 20--30\% & $0.130\pm0.003$ & $0.308\pm0.004$ & $0.27\pm0.02$    & 45 & 26\\
        &        & 30--40\% & $0.129\pm0.004$ & $0.307\pm0.006$ & $0.13\pm0.01$    & 59 & 26\\
        &        & 40--50\% & $0.128\pm0.005$ & $0.306\pm0.007$ & $0.060\pm0.006$  & 11 & 26\\
        &        & 50--60\% & $0.126\pm0.004$ & $0.300\pm0.006$ & $0.026\pm0.002$  & 15 & 25\\
        &        & 60--70\% & $0.124\pm0.003$ & $0.288\pm0.005$ & $0.011\pm0.001$  & 6  & 23\\
        &        & 70--80\% & $0.122\pm0.004$ & $0.264\pm0.011$ & $0.0050\pm0.0003$& 26 & 22\\
\cline{2-8}
        & $p$    & 0--5\%   & $0.136\pm0.005$ & $0.323\pm0.007$ & $7.74\pm1.00$    & 55 & 28\\
        &        & 5--10\%  & $0.135\pm0.005$ & $0.321\pm0.007$ & $2.90\pm0.25$    & 56 & 29\\
        &        & 10--20\% & $0.134\pm0.005$ & $0.318\pm0.006$ & $1.12\pm0.15$    & 40 & 29\\
        &        & 20--30\% & $0.132\pm0.004$ & $0.311\pm0.007$ & $0.50\pm0.03$    & 24 & 29\\
        &        & 30--40\% & $0.130\pm0.005$ & $0.308\pm0.007$ & $0.24\pm0.01$    & 13 & 29\\
        &        & 40--50\% & $0.128\pm0.003$ & $0.285\pm0.006$ & $0.12\pm0.01$    & 10 & 28\\
        &        & 50--60\% & $0.125\pm0.004$ & $0.274\pm0.008$ & $0.55\pm0.01$    & 13 & 28\\
        &        & 60--70\% & $0.123\pm0.004$ & $0.251\pm0.006$ & $0.024\pm0.004$  & 7  & 28\\
        &        & 70--80\% & $0.121\pm0.004$ & $0.231\pm0.007$ & $0.010\pm0.003$  & 23 & 29\\
\cline{1-8}
\end{tabular}%
\end{center}}
\end{table*}

\begin{table*}
{\small Table 1. Continued. \vspace{-.50cm}
\begin{center}
\begin{tabular}{cccccccc}\\ \hline\hline
Figure & Particle & Centrality & $T_0$ & $\beta_T$ & $N_0$ &
$\chi^2$ & dof\\ \hline
Fig. 3  &$\pi^+$ & 0--5\%   & $0.135\pm0.003$ & $0.320\pm0.005$ & $22.14\pm2.00$   & 4  & 28\\
Au-Au   &        & 5--10\%  & $0.133\pm0.003$ & $0.318\pm0.006$ & $8.50\pm2.00$    & 10 & 28\\
14.5 GeV&        & 10--20\% & $0.132\pm0.004$ & $0.317\pm0.006$ & $3.50\pm0.25$    & 12 & 28\\
        &        & 20--30\% & $0.131\pm0.004$ & $0.315\pm0.005$ & $1.60\pm0.13$    & 15 & 28\\
        &        & 30--40\% & $0.130\pm0.004$ & $0.314\pm0.007$ & $0.80\pm0.06$    & 10 & 28\\
        &        & 40--50\% & $0.128\pm0.004$ & $0.311\pm0.006$ & $0.40\pm0.03$    & 16 & 28\\
        &        & 50--60\% & $0.126\pm0.004$ & $0.307\pm0.007$ & $0.19\pm0.02$    & 7  & 28\\
        &        & 60--70\% & $0.124\pm0.004$ & $0.299\pm0.007$ & $0.093\pm0.014$  & 8  & 28\\
        &        & 70--80\% & $0.120\pm0.005$ & $0.291\pm0.005$ & $0.040\pm0.006$  & 14 & 28\\
\cline{2-8}
        & $K^+$  & 0--5\%   & $0.137\pm0.005$ & $0.318\pm0.007$ & $4.36\pm0.40$    & 16 & 26\\
        &        & 5--10\%  & $0.136\pm0.004$ & $0.314\pm0.008$ & $1.86\pm0.20$    & 8  & 26\\
        &        & 10--20\% & $0.135\pm0.005$ & $0.313\pm0.008$ & $0.70\pm0.08$    & 21 & 26\\
        &        & 20--30\% & $0.134\pm0.004$ & $0.312\pm0.006$ & $0.31\pm0.03$    & 15 & 26\\
        &        & 30--40\% & $0.132\pm0.004$ & $0.310\pm0.009$ & $0.14\pm0.01$    & 9  & 26\\
        &        & 40--50\% & $0.130\pm0.003$ & $0.308\pm0.008$ & $0.067\pm0.006$  & 7  & 24\\
        &        & 50--60\% & $0.128\pm0.006$ & $0.305\pm0.010$ & $0.025\pm0.003$  & 9  & 24\\
        &        & 60--70\% & $0.127\pm0.004$ & $0.294\pm0.007$ & $0.012\pm0.001$  & 3  & 22\\
        &        & 70--80\% & $0.125\pm0.005$ & $0.267\pm0.008$ & $0.0060\pm0.0006$& 4  & 20\\
\cline{2-8}
        & $p$    & 0--5\%   & $0.139\pm0.005$ & $0.335\pm0.009$ & $6.47\pm0.70$    & 22 & 25\\
        &        & 5--10\%  & $0.137\pm0.004$ & $0.328\pm0.008$ & $2.90\pm0.30$    & 20 & 25\\
        &        & 10--20\% & $0.135\pm0.003$ & $0.326\pm0.008$ & $1.03\pm0.12$    & 18 & 25\\
        &        & 20--30\% & $0.134\pm0.004$ & $0.323\pm0.007$ & $0.46\pm0.07$    & 15 & 25\\
        &        & 30--40\% & $0.132\pm0.004$ & $0.315\pm0.008$ & $0.21\pm0.03$    & 12 & 25\\
        &        & 40--50\% & $0.130\pm0.005$ & $0.311\pm0.007$ & $0.096\pm0.012$  & 11 & 25\\
        &        & 50--60\% & $0.128\pm0.005$ & $0.294\pm0.005$ & $0.042\pm0.007$  & 13 & 25\\
        &        & 60--70\% & $0.126\pm0.005$ & $0.270\pm0.008$ & $0.018\pm0.004$  & 18 & 25\\
        &        & 70--80\% & $0.123\pm0.004$ & $0.236\pm0.007$ & $0.0080\pm0.0019$& 26 & 25\\
\cline{1-8}
Fig. 4  &$\pi^+$ & 0--5\%   & $0.138\pm0.004$ & $0.322\pm0.004$ & $24.14\pm2.00$   & 9  & 26\\
Au-Au   &        & 5--10\%  & $0.137\pm0.004$ & $0.321\pm0.005$ & $9.50\pm0.80$    & 6  & 26\\
19.6 GeV&        & 10--20\% & $0.135\pm0.005$ & $0.319\pm0.008$ & $3.75\pm0.25$    & 7  & 26\\
        &        & 20--30\% & $0.134\pm0.004$ & $0.317\pm0.005$ & $1.97\pm0.16$    & 12 & 26\\
        &        & 30--40\% & $0.132\pm0.003$ & $0.315\pm0.005$ & $0.87\pm0.06$    & 18 & 26\\
        &        & 40--50\% & $0.130\pm0.004$ & $0.312\pm0.006$ & $0.43\pm0.04$    & 20 & 26\\
        &        & 50--60\% & $0.129\pm0.005$ & $0.311\pm0.008$ & $0.22\pm0.02$    & 16 & 26\\
        &        & 60--70\% & $0.128\pm0.004$ & $0.308\pm0.008$ & $0.10\pm0.01$    & 16 & 26\\
        &        & 70--80\% & $0.125\pm0.005$ & $0.303\pm0.009$ & $0.048\pm0.004$  & 14 & 26\\
\cline{2-8}
        & $K^+$  & 0--5\%   & $0.140\pm0.003$ & $0.320\pm0.007$ & $4.98\pm0.30$    & 24 & 26\\
        &        & 5--10\%  & $0.138\pm0.005$ & $0.319\pm0.009$ & $2.00\pm0.20$    & 33 & 26\\
        &        & 10--20\% & $0.136\pm0.005$ & $0.318\pm0.007$ & $0.75\pm0.05$    & 33 & 26\\
        &        & 20--30\% & $0.135\pm0.005$ & $0.314\pm0.007$ & $0.34\pm0.02$    & 21 & 26\\
        &        & 30--40\% & $0.133\pm0.004$ & $0.311\pm0.007$ & $0.16\pm0.03$    & 12 & 26\\
        &        & 40--50\% & $0.130\pm0.005$ & $0.309\pm0.009$ & $0.075\pm0.005$  & 15 & 26\\
        &        & 50--60\% & $0.129\pm0.004$ & $0.307\pm0.008$ & $0.036\pm0.005$  & 8  & 26\\
        &        & 60--70\% & $0.128\pm0.006$ & $0.300\pm0.010$ & $0.016\pm0.001$  & 23 & 26\\
        &        & 70--80\% & $0.126\pm0.004$ & $0.294\pm0.009$ & $0.0070\pm0.0003$& 25 & 26\\
\cline{2-8}
        & $p$    & 0--5\%   & $0.142\pm0.005$ & $0.338\pm0.006$ & $5.84\pm0.70$    & 41 & 29\\
        &        & 5--10\%  & $0.140\pm0.006$ & $0.336\pm0.008$ & $2.40\pm0.30$    & 28 & 25\\
        &        & 10--20\% & $0.138\pm0.005$ & $0.334\pm0.005$ & $0.91\pm0.12$    & 19 & 23\\
        &        & 20--30\% & $0.136\pm0.005$ & $0.324\pm0.007$ & $0.37\pm0.06$    & 41 & 23\\
        &        & 30--40\% & $0.133\pm0.004$ & $0.316\pm0.005$ & $0.18\pm0.03$    & 20 & 23\\
        &        & 40--50\% & $0.131\pm0.006$ & $0.312\pm0.008$ & $0.090\pm0.016$  & 8  & 23\\
        &        & 50--60\% & $0.129\pm0.005$ & $0.295\pm0.009$ & $0.042\pm0.007$  & 14 & 23\\
        &        & 60--70\% & $0.128\pm0.004$ & $0.275\pm0.008$ & $0.019\pm0.003$  & 2  & 23\\
        &        & 70--80\% & $0.125\pm0.004$ & $0.237\pm0.007$ & $0.0080\pm0.0013$& 11 & 23\\
\cline{1-8}
\end{tabular}%
\end{center}}
\end{table*}

\begin{table*}
{\small Table 1. Continued. \vspace{-.50cm}
\begin{center}
\begin{tabular}{cccccccc}\\ \hline\hline
Figure & Particle & Centrality & $T_0$ & $\beta_T$ & $N_0$ &
$\chi^2$ & dof\\ \hline
Fig. 5  &$\pi^+$ & 0--5\%   & $0.139\pm0.004$ & $0.326\pm0.006$ & $26.14\pm1.80$   & 5  & 26\\
Au-Au   &        & 5--10\%  & $0.138\pm0.004$ & $0.324\pm0.008$ & $11.07\pm2.00$   & 8  & 26\\
27 GeV  &        & 10--20\% & $0.136\pm0.005$ & $0.323\pm0.004$ & $4.25\pm0.25$    & 10 & 26\\
        &        & 20--30\% & $0.135\pm0.004$ & $0.322\pm0.004$ & $1.90\pm0.15$    & 14 & 26\\
        &        & 30--40\% & $0.133\pm0.003$ & $0.321\pm0.004$ & $0.98\pm0.10$    & 23 & 26\\
        &        & 40--50\% & $0.131\pm0.004$ & $0.320\pm0.006$ & $0.50\pm0.03$    & 26 & 26\\
        &        & 50--60\% & $0.130\pm0.005$ & $0.318\pm0.005$ & $0.23\pm0.03$    & 19 & 26\\
        &        & 60--70\% & $0.129\pm0.005$ & $0.317\pm0.009$ & $0.11\pm0.01$    & 21 & 26\\
        &        & 70--80\% & $0.128\pm0.004$ & $0.315\pm0.005$ & $0.048\pm0.005$  & 19 & 26\\
\cline{2-8}
        & $K^+$  & 0--5\%   & $0.142\pm0.005$ & $0.324\pm0.008$ & $5.11\pm0.60$    & 53 & 26\\
        &        & 5--10\%  & $0.140\pm0.005$ & $0.322\pm0.007$ & $2.15\pm0.20$    & 52 & 26\\
        &        & 10--20\% & $0.139\pm0.003$ & $0.321\pm0.008$ & $0.82\pm0.10$    & 55 & 26\\
        &        & 20--30\% & $0.137\pm0.006$ & $0.321\pm0.005$ & $0.37\pm0.03$    & 43 & 26\\
        &        & 30--40\% & $0.136\pm0.004$ & $0.320\pm0.007$ & $0.18\pm0.01$    & 25 & 26\\
        &        & 40--50\% & $0.134\pm0.004$ & $0.318\pm0.008$ & $0.86\pm0.01$    & 9  & 26\\
        &        & 50--60\% & $0.132\pm0.005$ & $0.316\pm0.005$ & $0.040\pm0.005$  & 8  & 26\\
        &        & 60--70\% & $0.130\pm0.004$ & $0.311\pm0.006$ & $0.014\pm0.003$  & 12 & 26\\
        &        & 70--80\% & $0.128\pm0.004$ & $0.304\pm0.007$ & $0.0070\pm0.0004$& 27 & 26\\
\cline{2-8}
        & $p$    & 0--5\%   & $0.144\pm0.004$ & $0.343\pm0.007$ & $5.31\pm0.40$    & 34 & 23\\
        &        & 5--10\%  & $0.143\pm0.004$ & $0.341\pm0.007$ & $2.20\pm0.22$    & 27 & 23\\
        &        & 10--20\% & $0.141\pm0.005$ & $0.336\pm0.007$ & $0.84\pm0.09$    & 21 & 23\\
        &        & 20--30\% & $0.139\pm0.005$ & $0.330\pm0.006$ & $0.37\pm0.05$    & 15 & 23\\
        &        & 30--40\% & $0.137\pm0.005$ & $0.326\pm0.008$ & $0.18\pm0.03$    & 9  & 23\\
        &        & 40--50\% & $0.134\pm0.005$ & $0.318\pm0.005$ & $0.090\pm0.016$  & 8  & 23\\
        &        & 50--60\% & $0.131\pm0.004$ & $0.300\pm0.008$ & $0.042\pm0.005$  & 3  & 23\\
        &        & 60--70\% & $0.129\pm0.004$ & $0.280\pm0.005$ & $0.019\pm0.002$  & 6  & 23\\
        &        & 70--80\% & $0.126\pm0.004$ & $0.257\pm0.007$ & $0.0070\pm0.0003$& 9  & 23\\
\cline{1-8}
Fig. 6  &$\pi^+$ & 0--5\%   & $0.141\pm0.004$ & $0.330\pm0.007$ & $27.84\pm2.30$   & 7  & 26\\
Au-Au   &        & 5--10\%  & $0.139\pm0.005$ & $0.328\pm0.007$ & $11.60\pm0.70$   & 14 & 26\\
39  GeV &        & 10--20\% & $0.138\pm0.004$ & $0.326\pm0.006$ & $4.50\pm0.30$    & 23 & 26\\
        &        & 20--30\% & $0.136\pm0.004$ & $0.325\pm0.005$ & $2.12\pm0.10$    & 38 & 26\\
        &        & 30--40\% & $0.135\pm0.003$ & $0.324\pm0.008$ & $1.05\pm0.08$    & 42 & 26\\
        &        & 40--50\% & $0.135\pm0.005$ & $0.322\pm0.005$ & $0.52\pm0.02$    & 36 & 26\\
        &        & 50--60\% & $0.134\pm0.004$ & $0.321\pm0.008$ & $0.27\pm0.02$    & 39 & 26\\
        &        & 60--70\% & $0.132\pm0.004$ & $0.320\pm0.007$ & $0.12\pm0.01$    & 36 & 26\\
        &        & 70--80\% & $0.130\pm0.005$ & $0.319\pm0.008$ & $0.062\pm0.005$  & 51 & 26\\
\cline{2-8}
        & $K^+$  & 0--5\%   & $0.148\pm0.004$ & $0.328\pm0.005$ & $5.29\pm0.40$    & 35 & 26\\
        &        & 5--10\%  & $0.147\pm0.004$ & $0.327\pm0.006$ & $2.30\pm0.15$    & 15 & 26\\
        &        & 10--20\% & $0.146\pm0.005$ & $0.328\pm0.005$ & $0.90\pm0.08$    & 29 & 26\\
        &        & 20--30\% & $0.145\pm0.006$ & $0.324\pm0.009$ & $0.40\pm0.03$    & 19 & 26\\
        &        & 30--40\% & $0.144\pm0.005$ & $0.323\pm0.008$ & $0.19\pm0.01$    & 12 & 26\\
        &        & 40--50\% & $0.143\pm0.005$ & $0.321\pm0.006$ & $0.090\pm0.010$  & 10 & 26\\
        &        & 50--60\% & $0.142\pm0.003$ & $0.317\pm0.006$ & $0.0040\pm0.0004$& 12 & 26\\
        &        & 60--70\% & $0.140\pm0.004$ & $0.316\pm0.005$ & $0.019\pm0.001$  & 15 & 26\\
        &        & 70--80\% & $0.138\pm0.005$ & $0.313\pm0.008$ & $0.0083\pm0.0003$& 18 & 26\\
\cline{2-8}
        & $p$    & 0--5\%   & $0.149\pm0.005$ & $0.359\pm0.008$ & $4.38\pm0.50$    & 34 & 22\\
        &        & 5--10\%  & $0.148\pm0.004$ & $0.348\pm0.006$ & $1.94\pm0.30$    & 36 & 22\\
        &        & 10--20\% & $0.146\pm0.005$ & $0.346\pm0.006$ & $0.80\pm0.12$    & 22 & 22\\
        &        & 20--30\% & $0.145\pm0.004$ & $0.340\pm0.007$ & $0.33\pm0.05$    & 16 & 22\\
        &        & 30--40\% & $0.144\pm0.004$ & $0.335\pm0.005$ & $0.16\pm0.03$    & 8  & 22\\
        &        & 40--50\% & $0.144\pm0.004$ & $0.330\pm0.006$ & $0.078\pm0.014$  & 13 & 22\\
        &        & 50--60\% & $0.143\pm0.004$ & $0.300\pm0.006$ & $0.040\pm0.005$  & 1  & 22\\
        &        & 60--70\% & $0.139\pm0.004$ & $0.281\pm0.005$ & $0.017\pm0.002$  & 4  & 22\\
        &        & 70--80\% & $0.127\pm0.004$ & $0.274\pm0.007$ & $0.0080\pm0.0006$& 10 & 22\\
\cline{1-8}
\end{tabular}%
\end{center}}
\end{table*}

\begin{figure*}[htb!]
\begin{center}
\hskip-0.153cm
\includegraphics[width=15cm]{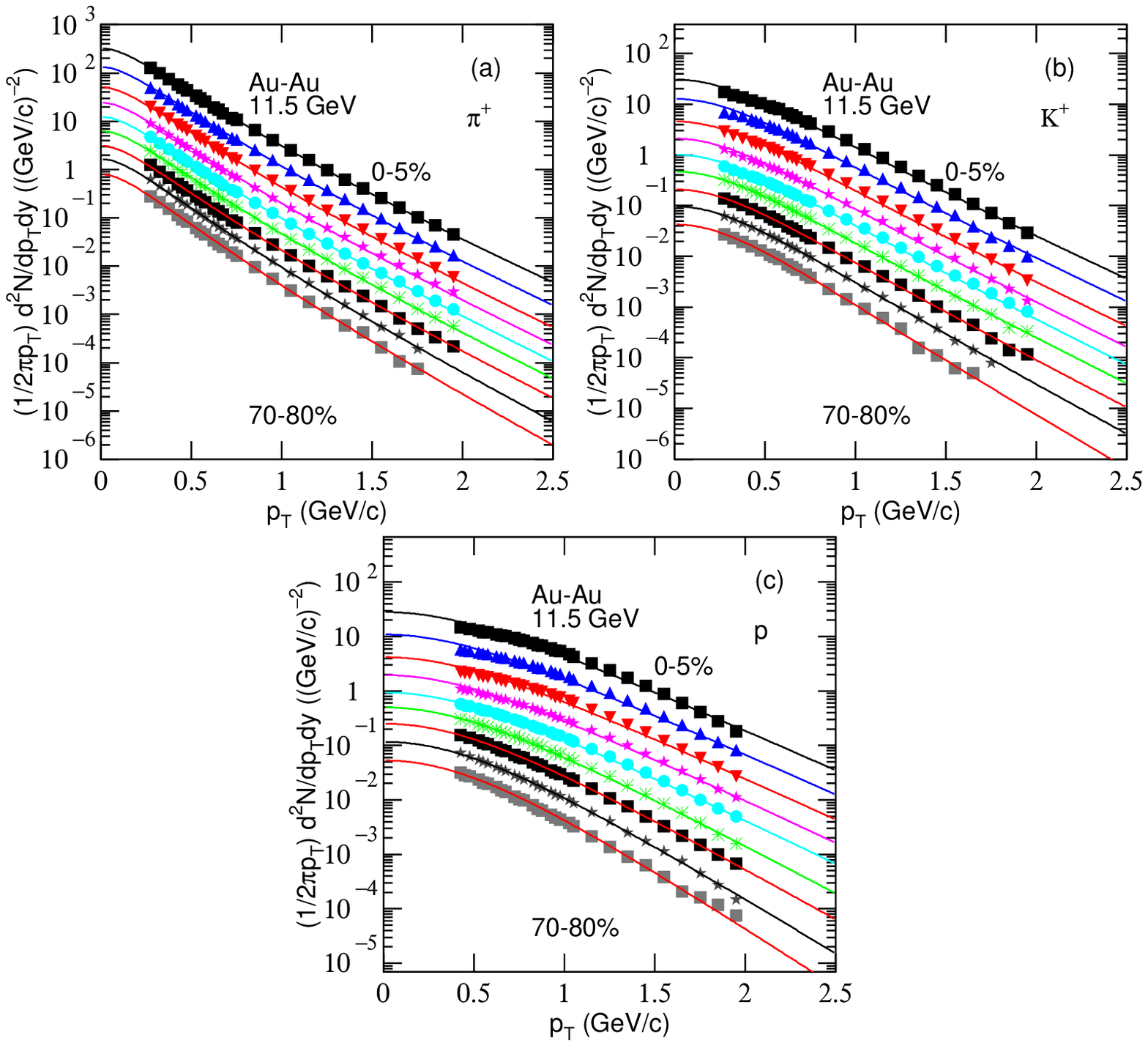}
\end{center}
Fig. 2. The same as Fig. 1, but showing the results at
$\sqrt{s_{NN}}=11.5$ GeV.
\end{figure*}

Figure 2 is the same as Fig. 1, but it shows the $p_T$ spectra at
$\sqrt{s_{NN}}=11.5$ GeV. One can see that Eq. (1) fits well the
data in Au-Au collisions at 11.5 GeV at the RHIC-BES.

Figure 3 is also the same as Fig. 1, but it shows the $p_T$
spectra at $\sqrt{s_{NN}}=14.5$ GeV, where the data are cited from
ref.~\cite{28}. Once again, Eq. (1) fits well the data in Au-Au
collisions at 14.5 GeV at the RHIC-BES.

Figures 4--6 are also the same as Fig. 1, but they show the $p_T$
spectra at $\sqrt{s_{NN}}=19.6$, 27 and 39 GeV, respectively. Once
more, Eq. (1) fits well the data in Au-Au collisions at other
RHIC-BES energies.

\begin{figure*}[htb!]
\begin{center}
\hskip-0.153cm
\includegraphics[width=15cm]{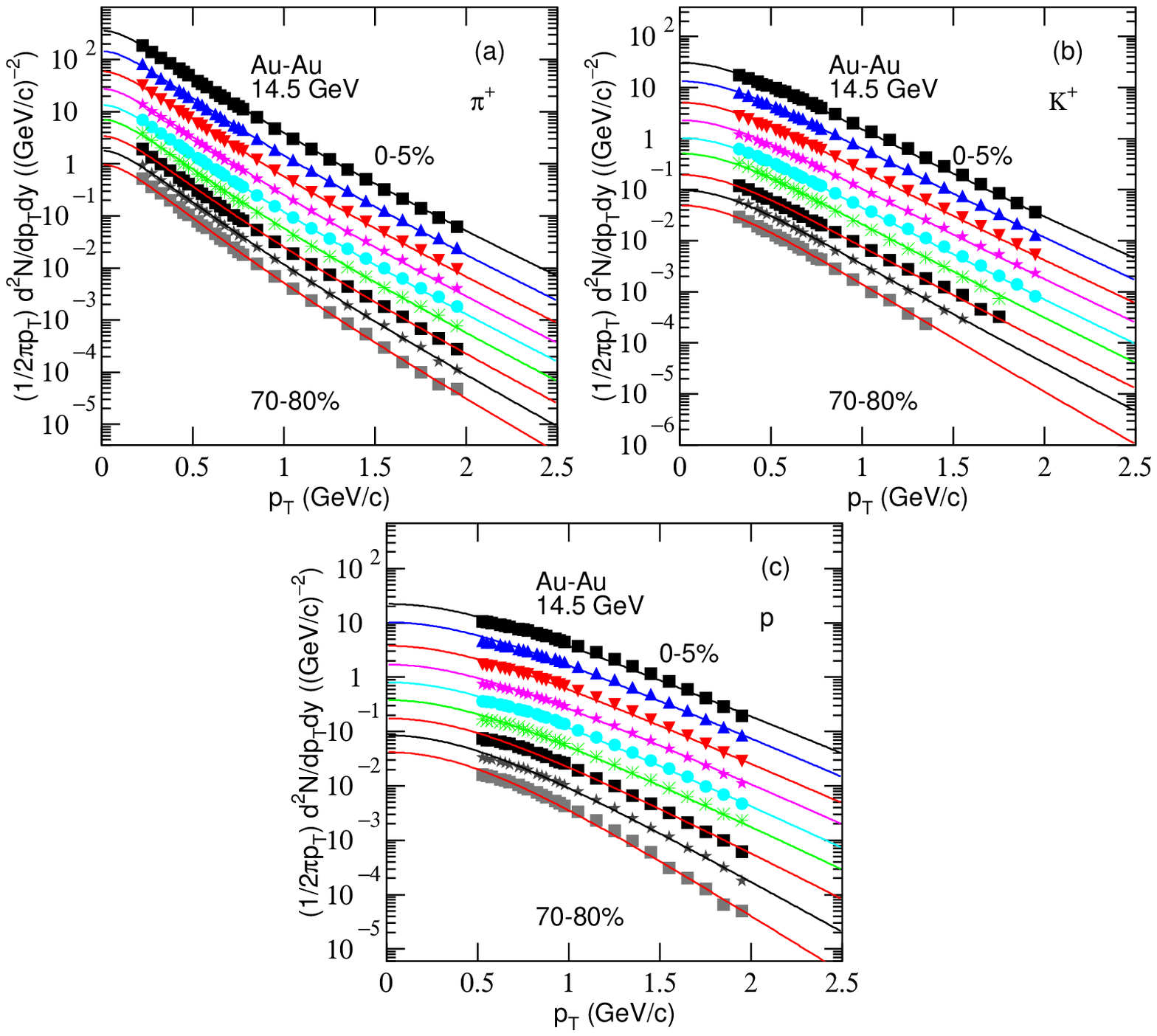}
\end{center}
Fig. 3. The same as Fig. 1, but showing the results at
$\sqrt{s_{NN}}=14.5$ GeV, where the data are cited from
ref.~\cite{28}.
\end{figure*}

\begin{figure*}[htb!]
\begin{center}
\hskip-0.153cm
\includegraphics[width=15cm]{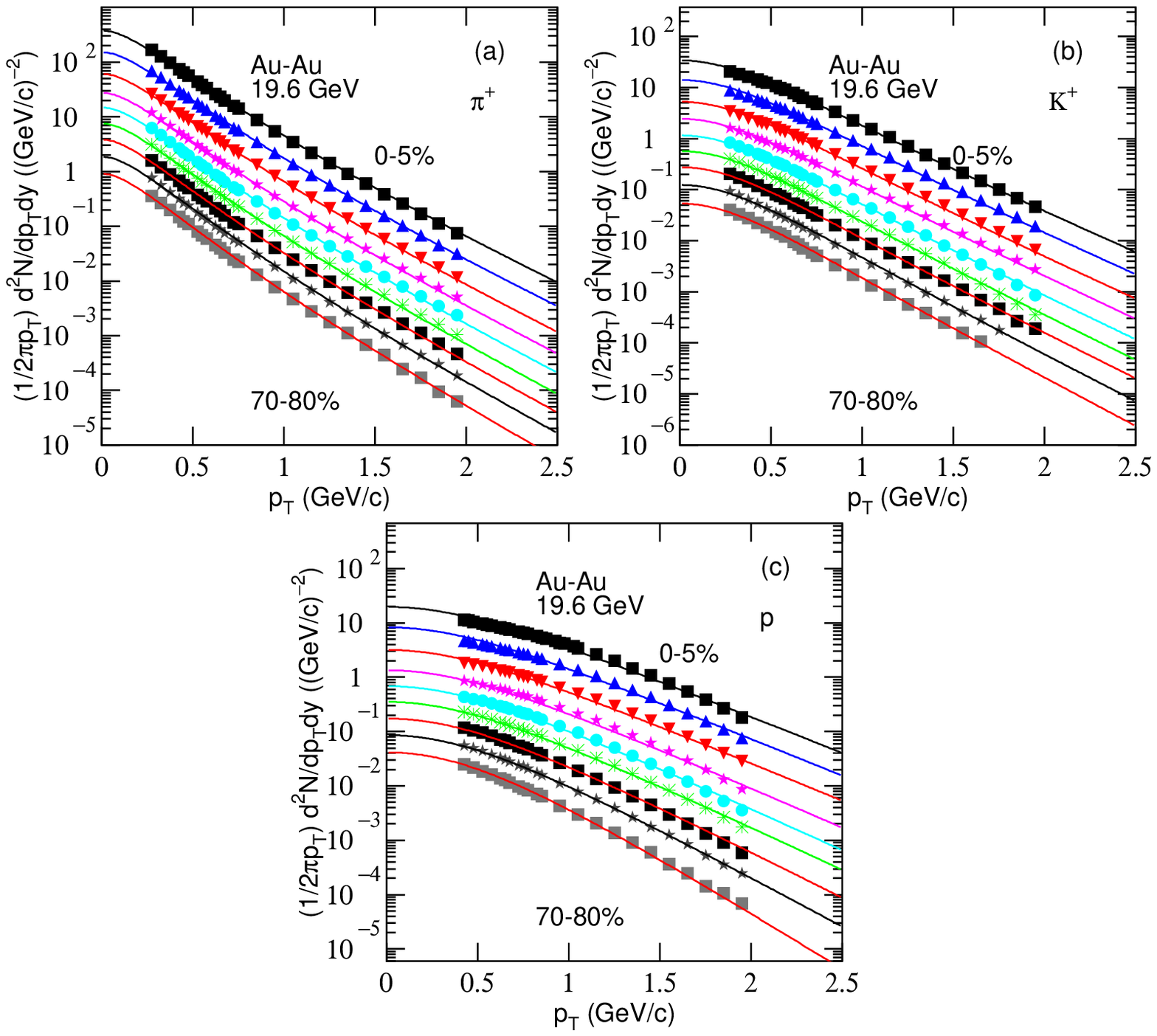}
\end{center}
Fig. 4. The same as Fig. 1, but showing the results at
$\sqrt{s_{NN}}=19.6$ GeV.
\end{figure*}

\begin{figure*}[htb!]
\begin{center}
\hskip-0.153cm
\includegraphics[width=15cm]{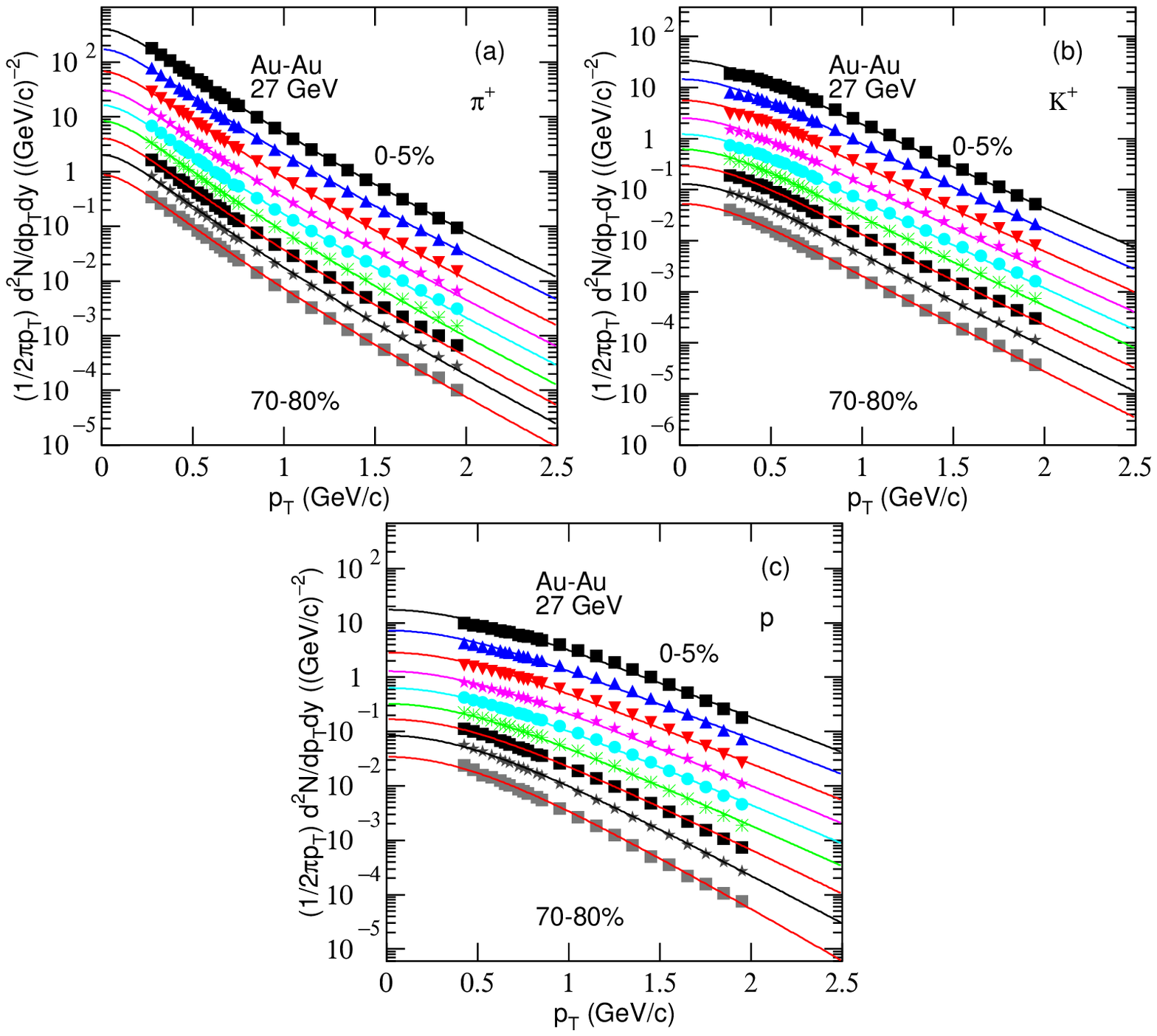}
\end{center}
Fig. 5. The same as Fig. 1, but showing the results at
$\sqrt{s_{NN}}=27$ GeV.
\end{figure*}

\begin{figure*}[htb!]
\begin{center}
\hskip-0.153cm
\includegraphics[width=15cm]{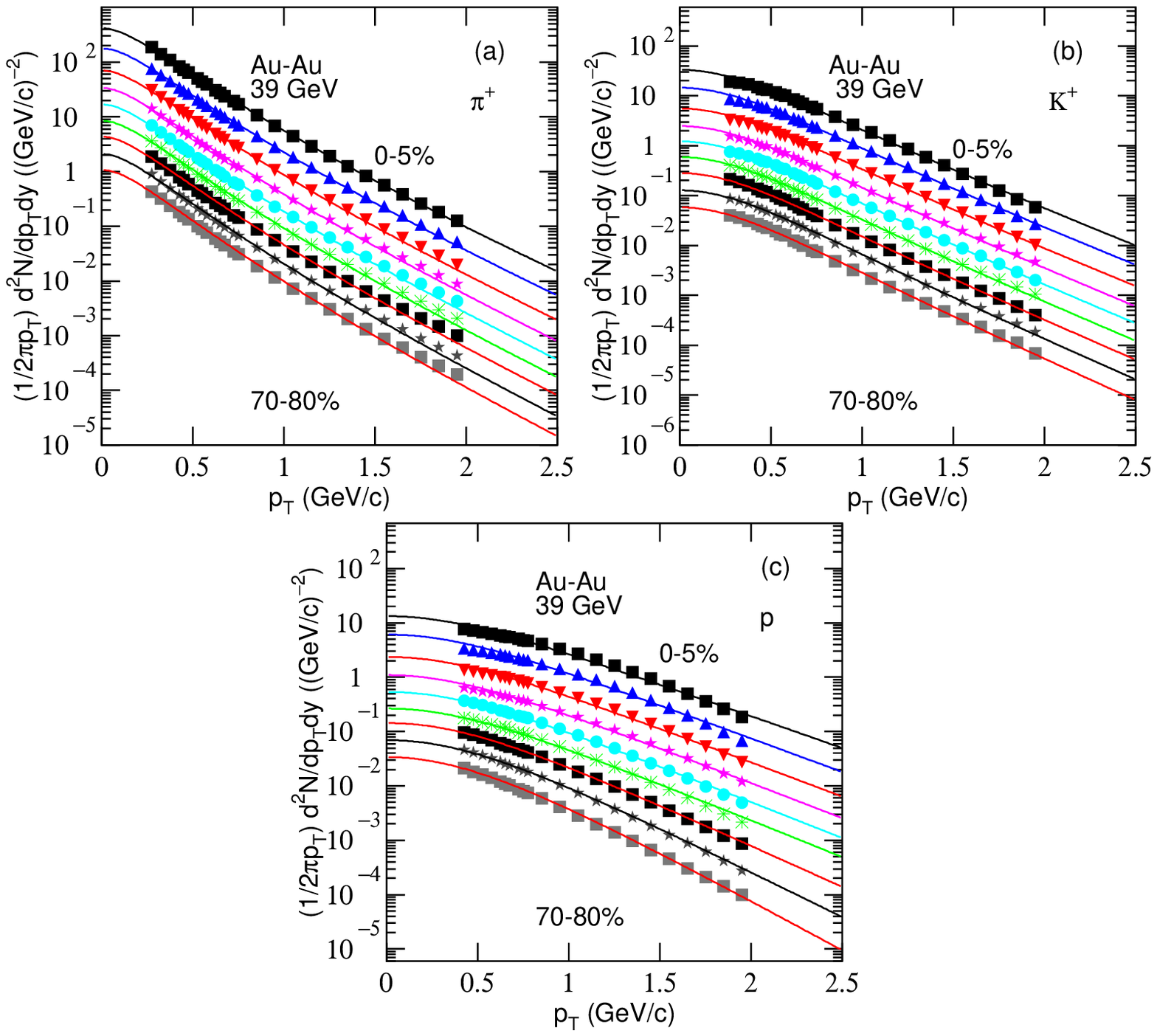}
\end{center}
Fig. 6. The same as Fig. 1, but showing the results at
$\sqrt{s_{NN}}=39$ GeV.
\end{figure*}

\begin{figure*}[htb!]
\begin{center}
\includegraphics[width=14.cm]{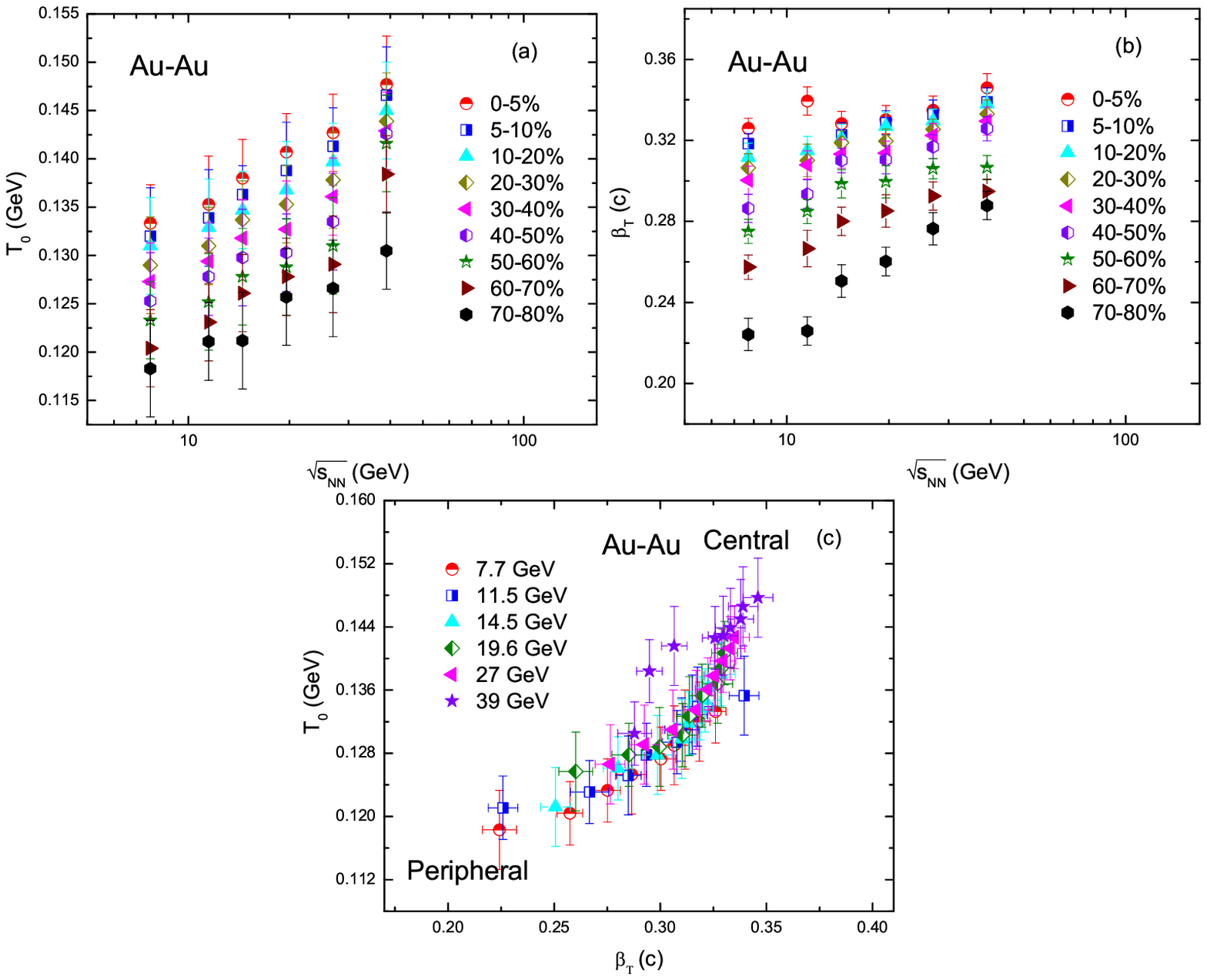}
\end{center}
Fig. 7. Dependences of weight averages (a) $T_0$ and (b) $\beta_T$
on $\sqrt{s_{NN}}$ for different event centralities as well as (c)
$T_0$ on $\beta_T$ for different collision energies and event
centralities. The different symbols display different centrality
classes in Figs. 7(a) and 7(b) or different collision energies in
Fig. 7(c), which are averaged by weighting the yields of different
particles which are listed in Table 1.
\end{figure*}

\begin{figure*}[htb!]
\begin{center}
\includegraphics[width=14.cm]{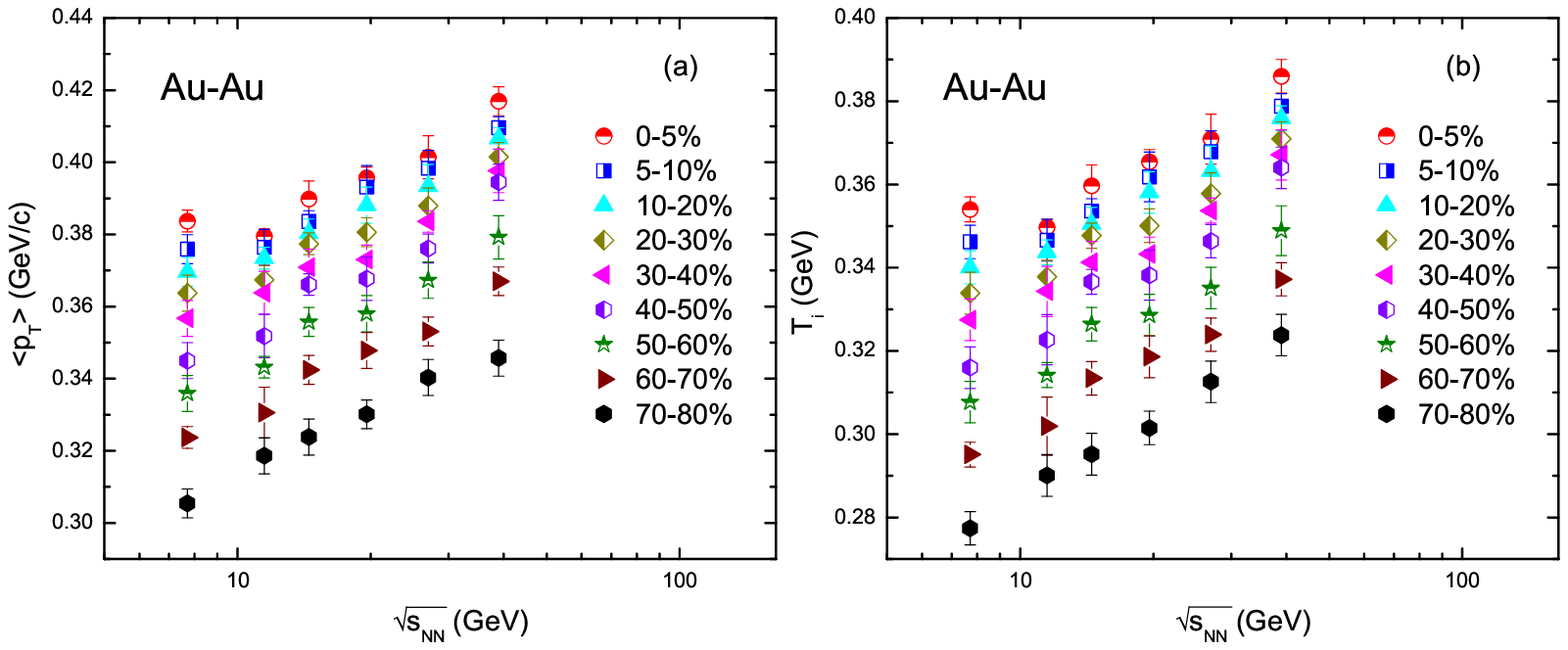}
\end{center}
Fig. 8. Dependences of weight averages (a) $\langle p_T\rangle$
and (b) $T_i$ on $\sqrt{s_{NN}}$ for different event centralities.
The different symbols display different centrality classes, which
are averaged by weighting the yields of different particles which
are listed in Table 1.
\end{figure*}

It is noteworthy to point out that Eq. (1) for the blast-wave
model in the system is assumed to be in local thermodynamic
equilibrium and therefore, a single $T_0$ and $\beta_T$ should be
obtained by the weight average of different particles species. To
see clearly the trends of weight average parameters, Figs. 7(a)
and 7(b) show the dependences of weight averages $T_0$ and
$\beta_T$ on $\sqrt{s_{NN}}$ for different event centralities. The
symbols represent the parameter values averaged by weighting the
yields of different particles which are listed in Table 1. One can
see that $T_0$ and $\beta_T$ increase with the increase of
$\sqrt{s_{NN}}$ from 7.7 to 39 GeV. Meanwhile, $T_0$ and $\beta_T$
increase with the increase of event centrality from periphery to
center.

In addition, the variation of weight averages $T_0$ on $\beta_T$
for different collision energies and event centralities are
displayed in Fig. 7(c), where the symbols represent the parameter
values averaged by weighting the yields of different particles.
One can see that $T_0$ increases with the increase of $\beta_T$.
At higher energy and in central collisions, one see larger $T_0$
and $\beta_T$. There is a positive correlation between $T_0$ and
$\beta_T$.

The dependences of mean transverse momentum ($\langle p_T\rangle$)
and initial temperature ($T_i=\sqrt{\langle p_T^2
\rangle/2}$~\cite{20a,20b,20c}) on $\sqrt{s_{NN}}$ for different
event centralities obtained by weighting the yields of different
particles are shown in Figs. 8(a) and 8(b) respectively. One can
see that $\langle p_T\rangle$ and $T_i$ increase with the increase
of $\sqrt{s_{NN}}$ from 7.7 to 39 GeV. Meanwhile, $\langle
p_T\rangle$ and $T_i$ increase with the increase of event
centrality from periphery to center.

The reason for increasing of $T_0$ and $\beta_T$ with the increase
of collision energy is due to the fact that more energies are
deposited in collisions at higher energy in the considered
RHIC-BES energy range. Meanwhile, the system size at higher energy
decreases due to relativistic constriction effect, which results
in a smaller volume, then a larger energy density and larger
$T_0$. Meanwhile, at higher energy, the squeeze is more violent,
which results in a rapider expansion and larger $\beta_T$.

The reason for increasing of $T_0$ and $\beta_T$ with the increase
of event centrality is due to the fact that the central collisions
contain more nucleons than the peripheral collisions, then more
energies are deposited in central collisions. Meanwhile a rapider
expansion appears due to more violent squeeze in central
collisions, comparatively to peripheral collisions. As a result,
$T_0$ and $\beta_T$ in central collisions are larger than those in
peripheral collisions.

Because of $\langle p_T\rangle$ and $T_i$ being positive
correlation with $T_0$ and $\beta_T$, the increasing of $T_0$ and
$\beta_T$ with the increases of collision energy and event
centrality result naturally in the increasing of $\langle
p_T\rangle$ and $T_i$ with the increases of collision energy and
event centrality. This work shows that the two free parameters
$T_0$ and $\beta_T$ and the two derived parameters $\langle
p_T\rangle$ and $T_i$ appear similar law on the dependences of
collision energy and event centrality. In particular, $\langle
p_T\rangle$ and $T_i$ are model-independent, though we obtain them
from model-dependent free parameters $T_0$ and $\beta_T$ in this
work. In fact, $\langle p_T\rangle$ and $T_i$ can be obtained by
the $p_T$ data themselves if the data are across the possible
$p_T$ range.

It should be noted that there is entanglement in the extraction of
$T_0$ and $\beta_T$. In fact, if one uses a smaller $T_0$ and a
larger $\beta_T$ for central collisions, a decreasing trend for
$T_0$ from peripheral to central collisions can be obtained.
Meanwhile, a negative correlation between $T_0$ and $\beta_T$ can
also be obtained. Thus, this situation is in agreement with some
current references~\cite{26,48,49}. If one even uses an almost
invariant or slightly larger $T_0$ and a properly larger $\beta_T$
for central collisions, an almost invariant or slightly increase
trend for $T_0$ from peripheral to central collisions can be
obtained~\cite{49a}. To show the flexibility in the extraction of
$T_0$ and $\beta_T$, this work has reported an increasing trend
for $T_0$ from peripheral to central collisions, and a positive
correlation between $T_0$ and $\beta_T$.

This whole phenomenal analysis results in degree of thermal motion
and collective expansion, that are reflected by $T_0$ and
$\beta_T$. With the increasing collision energy, the system may
undergo different evolution processes. In the considered RHIC-BES
energy range, the violent degree of collisions increase with
increasing the collision energy. The trends of $T_0$ and $\beta_T$
show approximately monotonous increase in which large fluctuation
does not appear. The evolution processes at the considered six
energies show similar behaviors to each other.
\\

{\section{Conclusions}}

The main observations and conclusions are summarized here.

(a) Based on the data-driven analysis, the blast-wave model with
Boltzmann-Gibbs statistics is used to analyze the collision energy
dependent and event centrality dependent double-differential
transverse momentum spectra of charged particles ($\pi^+$, $K^+$
and $p$) produced in the mid-rapidity interval in Au-Au collisions
at the RHIC-BES energies. The contribution of soft excitation is
considered in this work, but the contribution of hard process is
not excluded if available.

(b) As the free parameters, the kinetic freeze-out temperature
$T_0$ and transverse flow velocity $\beta_T$ are extracted by the
blast-wave model. Both $T_0$ and $\beta_T$ increase with the
increase of collision energy due to more violent collisions at
higher energy. The two parameters also increase with the increase
of centrality, as the central collisions contain more nucleons
which means more energy deposited and more violent collisions and
squeeze, comparing with peripheral collisions.

(c) As the derived parameters, the mean transverse momentum
$\langle p_T\rangle$ and initial temperature $T_i$ appear similar
law to the free parameters $T_0$ and $\beta_T$ when we study the
dependences of parameters on collision energy and event
centrality. Although $T_0$ and $\beta_T$ are model-dependent,
$\langle p_T\rangle$ and $T_i$ are generally model-independent.
There is no large fluctuation in the excitation function of the
considered parameters at the RHIC-BES, which means similar
collision mechanism.
\\
\\

{\bf Acknowledgments}

We thank Dr. Muhammad Usman Ashraf for his kind help. This work
was supported by the National Natural Science Foundation of China
under Grant No. 11575103, the Chinese Government Scholarship
(China Scholarship Council), the Scientific and Technological
Innovation Programs of Higher Education Institutions in Shanxi
(STIP) under Grant No. 201802017, the Shanxi Provincial Natural
Science Foundation under Grant No. 201701D121005, and the Fund for
Shanxi ``1331 Project" Key Subjects Construction.
\\
\\
{\bf Data availability}

The data used to support the findings of this study are included
within the article and are cited at relevant places within the
text as references.
\\
\\
{\bf Compliance with Ethical Standards}

The authors declare that they are in compliance with ethical
standards regarding the content of this paper.
\\
\\
{\bf Conflict of Interest}

The authors declare that there are no conflicts of interest
regarding the publication of this paper. The funders had no role
in the design of the study; in the collection, analyses, or
interpretation of the data; in the writing of the manuscript, or
in the decision to publish the results.
\\
\\

\end{multicols}
\end{document}